# Superior bit error rate and jitter due to improved switching field distribution in exchange spring magnetic recording

D. Suess, M. Fuger, C. Abert, F. Bruckner, R. Windl, P. Palmesi, A. Buder, C. Vogler

Christian Doppler Laboratory for Advanced Magnetic Sensing and Materials,
Institute for Solid State Physics, TU Wien, Wiedner Hauptstrasse 8-10, 1040 Vienna.

[a]Electronic mail: dieter.suess@tuwien.ac.at

*Index terms*-- exchange spring media, switching field distribution, high-density recording

**Abstract:** We report two effects that lead to a reduction of the switching field distribution in exchange spring media. The first effect relies on a subtle mechanism of the interplay between exchange coupling between soft and hard layers and anisotropy that allows significant reduction of the switching field distribution in exchange spring media. This effect reduces the switching field distribution by about 30% compared to single-phase media. A second effect is that due to the improved thermal stability of exchange spring media over single-phase media, thermal fluctuation leads to reduced fundamental transition jitter. The influence of this overall improved switching field distribution on the transition jitter in granular recording and the bit error rate in bit-patterned magnetic recording is discussed. The transition jitter in granular recording for a distribution of $K_1$ values of 3% in the hard layer, taking into account thermal fluctuations during recording, is estimated to be $a$ = 0.78 nm, which is similar to the best reported calculated jitter in optimized heat-assisted recording media.

# I. Introduction

Magnetic recording at high density requires small magnetic grains or islands and narrow distribution of the material properties. In addition, intrinsic noise due to the thermal fluctuations during writing has to be minimized. This is a particular challenge in heat-assisted recording, where the write process occurs at elevated temperature [1],[2]. In heat-assisted recording for smaller grain sizes, both (i) the fundamental jitter due to thermal fluctuations during writing as well as (ii) the $T_c$ distributions are expected to increase.
In order to break the dilemma of designing a media that has a good writability and good thermal stability for non-heat-assisted recording, exchange spring media were proposed in 2004 [3]. The original goal of exchange spring media was (i) to allow for a media that is stable, even for smaller grain sizes, and (ii) to obtain an improved switching field distribution (SFD) due to the Kondorsky dependence of the switching field [4] as a function of field angle [3, 5].
Since about 2007, exchange spring media have been used in current products of hard disk drives. Interestingly, the grain size in these media has not decreased. Nevertheless, exchange spring media lead to an improved signal-to-noise ratio. This can obviously not attribute to the main original goal of a small grain size. We believe the reason can be found in an improved switching field distribution in exchange spring media, which was predicted by micromagnetic simulations [6] . Experimentally, the improved switching field distribution was reported by various groups [7, 8].
The switching field distribution is essential in order to improve the areal density in both bit-patterned magnetic recording and magnetic recording on granular films.

# II. Switching field distribution due to variation of anisotropy for fully coupled exchange spring grains

In this section, we review findings on the switching field distribution for exchange spring media and graded media in cases in which the layers with different anisotropies are fully exchange-coupled.

### A. *Bilayer exchange spring media*

In order to study in detail the improved error rate in exchange spring media, we investigate the influence of the switching field as a function of the anisotropy in the media.

In single-phase media, the switching field at $T = 0K$ is determined according to the Stoner-Wohlfarth theory as:

$$\mu_0 H_{s,0} = \mu_0 \frac{2K_{hard}}{J_{s,hard}}. \tag{1}$$

Here, $K_{hard}$ is the anisotropy constant in the media, and $J_{s,hard}$ is the magnetic polarization. In the case of the single-domain particle, the switching field changes linearly with the anisotropy of the media:

$$\frac{\sigma_{\mu_0 H_{s,0}}}{\mu_0 H_{s,0}} = \frac{\Delta \mu_0 H_{s,0}}{\mu_0 H_{s,0}} = \frac{\mu_0 H_{s,grain,SW}(K_1 + \Delta K_1) - \mu_0 H_{s,grain,SW}(K_1)}{\mu_0 H_{s,grain,SW}(K_1)} = \frac{\Delta K_1}{K_1}. \tag{2}$$

For the case of the exchange spring media, the switching field as a function of hard- or soft-layer anisotropy becomes a non-monotonic function due to the interplay between pinning field and nucleation field.

Hence, the estimate of the switching field distribution requires numerical simulations. In the work of Suess et al. [6], the switching field distribution of exchange spring media is investigated for uncorrelated and correlated distributions of the anisotropy constant in the soft layer and hard layer. A constant mean value of the anisotropy in the hard layer is assumed. The switching field distribution is calculated for various mean values of the soft-layer anisotropy. A full coupling between the soft layer and hard layer is assumed. In Ref [6], it is concluded that for a value of the anisotropy constant in the soft layer, which is 1/5 of the anisotropy of the hard layer, the minimum switching field distribution is obtained. Due to this minimum, small variations of the anisotropy value of the soft layer or, equivalently, variations of the anisotropy constant of the hard layer do not change the switching field in the first order. Hence, the switching field distribution is significantly improved. For the case that the soft-layer anisotropy and hard-layer anisotropy are perfectly correlated, an improvement of the switching field distribution (SFD) by 35% is obtained. For the case that the layers are decoupled, the SFD is improved by 50% [6].

*B. Graded media with multiple exchange coupled layers*

In the case of a graded media where the anisotropy constant increases quadratically as a function of the depth of the magnetic media, the switching field no longer depends linearly as a function of the anisotropy of the hardest layer. Using the equation as derived in Ref [9], the switching field at $T = 0K$ is given by:

$$\mu_0 H_{s,0,graded} = \mu_0 2/J_s \sqrt{AK_{1,max}}/t. \tag{3}$$

Here, $A$ is the exchange constant, and $t$ the layer thickness. Interestingly, if all layers are correlated, the switching field only varies according to $\sqrt{K_{max}}$. Hence, a distribution of $\Delta K_{1,max}$ only leads to:

$$\frac{\sigma_{\mu_0 H_{s,0}}}{\mu_0 H_{s,0}} = \frac{\Delta H}{H} = \frac{1}{2}\frac{\Delta K_{1,max}}{K_{1,max}}. \tag{4}$$

If the anisotropy of each layer of a graded media is uncorrelated and shows a distribution of 10%, the switching field distribution is given in Ref [6] to be only 3%. Hence, it is even smaller as for the correlated case, as given by Eq. (3). Hence, in graded media, a distribution of 10% of the $K_1$ values leads at most to a switching field distribution of 5%!

## III. Optimizing switching field distribution by varying exchange between hard and soft layers

In contrast to Ref [6], where the exchange coupling between the soft and hard layer was kept constant and the anisotropy in the soft layer was changed, in this chapter, we investigate the switching field distribution for exchange spring media as a function of the coupling strength. Whereas full coupling will be beneficial for sufficient thick exchange spring media, decoupling the layers will be beneficial for very thin layers (thickness smaller than 10 nm), which is important, for example, for bit-patterned media.

In order to study the influence of the anisotropy on the switching field for the grain of a recording media – or, alternatively, interpreted one island of a patterned media - we apply a field pulse for a time of 1 ns. The rise time from zero field to the maximum field is 0.1 ns. The field is applied at an angle of 10° with respect to the easy axis. The minimum field pulse strength that is required in order to switch the element is denoted as $\mu_0 H_{s,0}$. The subscript zero in the switching field denotes that in this chapter, all simulations are performed without thermal fluctuations. Hence, the simulations represent simulation at $T = 0$ K.

For all investigated media, the thickness of the soft layer equals the thickness of the hard layer. In all media, the total layer thickness is 10.4 nm, the magnetic polarization $J_s = 0.753$ T, and the exchange constant within a grain is $A = 10$ pJ/m. The damping constant is $\alpha = 0.02$. For the exchange spring structure, the soft-layer anisotropy constant is 1/5 of the hard-layer anisotropy constant. The exchange coupling between the soft layer and hard layer is varied. The simulated geometry is shown in Figure 1.

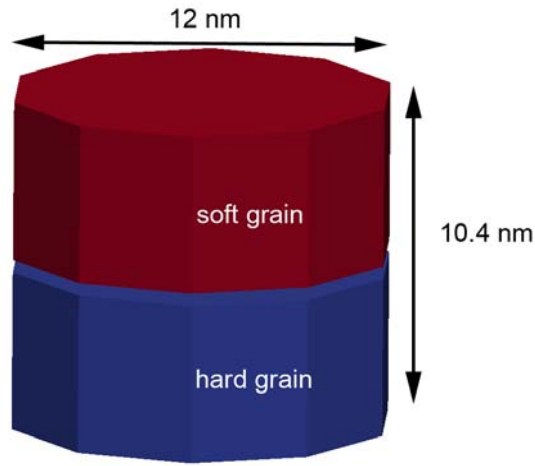

Figure 1: Geometry of the simulated bit-patterned island. The diameter is 12 nm, and the height is 10.4 nm. For the case of single-phase media, the blue and red grains are fully coupled. For the exchange spring structure, the coupling strength is varied.

The switching field as a function of the media layer anisotropy is plotted in Figure 2 . The procedure as follows is used in order to extract the switching field distribution:
   (i) We extract from Figure 2 the anisotropy of the hardest layer ($K_{min}$), which leads to a switching field of $\mu_0 H_{s,0,min} = 1$ T.
   (ii) We extract from Figure 2 the anisotropy of the hardest layer ($K_{max}$), which leads to a switching field of $\mu_0 H_{s,0,max} = 1$ T x 1.18.

We define the switching field variation as:

$$\sigma_r = \sigma_{\mu_0 H_{s,0}} / \mu_0 H_{s,0} = \frac{H_{s,0,max} - H_{s,0,min}}{H_{s,0,aver}} = 0.18. \quad (5)$$

$$H_{s,0,aver} = \frac{H_{s,0,max} + H_{s,0,min}}{2}. \quad (6)$$

Using the procedure mentioned above, we obtain for a given switching field distribution the according anisotropy distribution:

$$\sigma_k = \frac{K_{max} - K_{min}}{K_{aver}}. \quad (7)$$

The values are summarized in Table 2. From Table 2, it can be seen that a larger variation of the anisotropy constant can be tolerated in exchange spring media than in the single-phase media, which leads to the same switching field distribution.

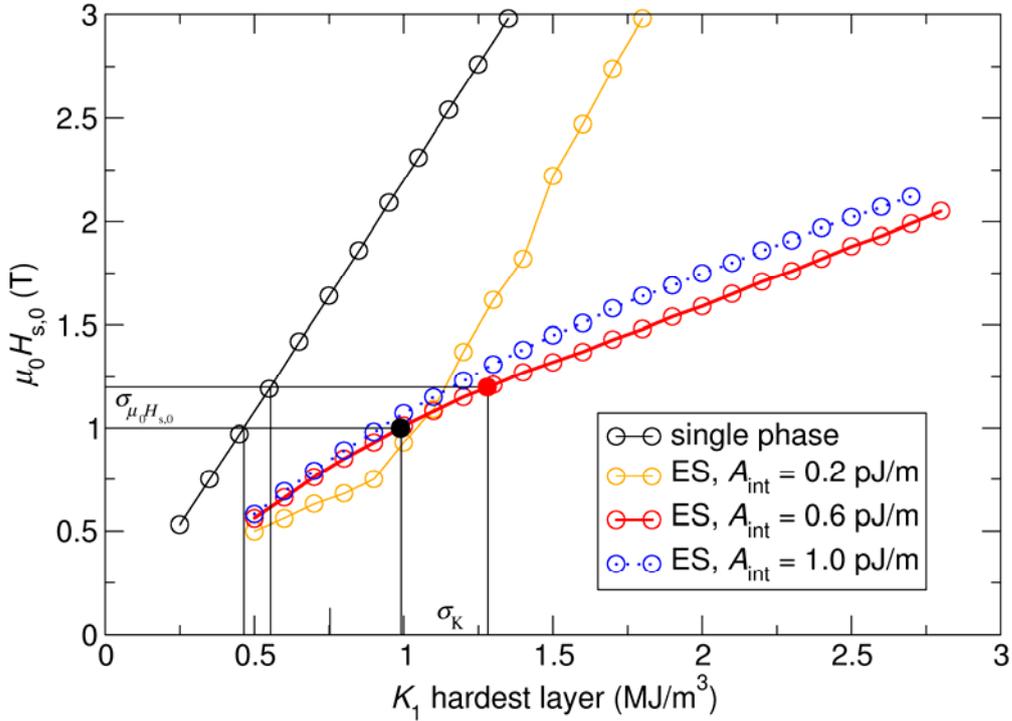

Figure 2: Field pulse strength, $\mu_0 H_{s,0}$(T), that is required to switch different media designs. The anisotropy of the media is changed.

Table 1: Dependence of switching field distributions due to variation of the anisotropy constant for $A_{int}$ = 0.6 pJ/m. It is assumed that the intrinsic anisotropy distribution is $\sigma_K$ =3%.

|  | Single-Phase | Exchange Spring |
|---|---|---|
| $\sigma_{\mu_0 H_{c,0}}$ (T) | 0.031 | 0.02 |
| $\sigma_r / \sigma_K$ | 1.05 | 0.69 |

In the following, we aim to understand the origin why a variation of the anisotropy values of the soft layer and hard layer in exchange spring media leads to a smaller change of the switching field compared to single-phase media. We investigate two different exchange spring media with different couplings between the soft and the hard layer.

A. *Intermediate coupled case ($A_{int}$ = 0.6 pJ/m):*

We start with the exchange spring media design with $A_{int}$ = 0.6 pJ/m, as shown by the red line in Figure 2. The open black dot and filled red dot represent an exchange spring media design with an anisotropy value in the hard layer of $K_{1,hard}$ = 1 MJ/m³ and $K_{1,hard}$ = 1.28 MJ/m³, respectively.

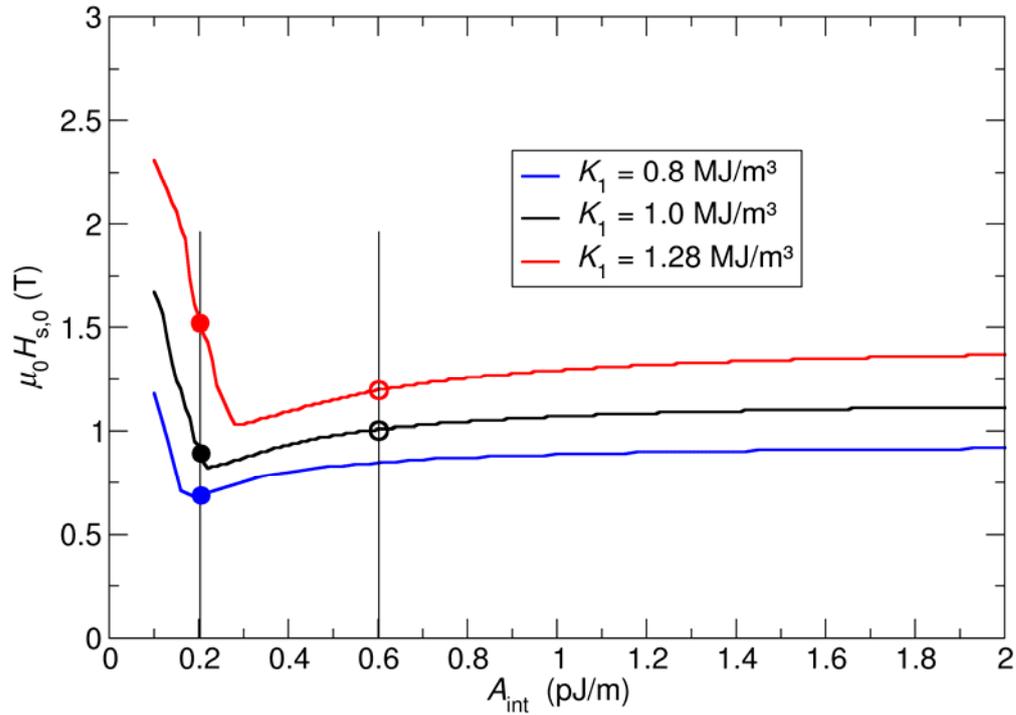

Figure 3: Required field pulse strength as a function of the exchange coupling between the layers for three different values of the anisotropy of the exchange spring media.

It is important to note that the minimum of the switching pulse strength is shifted to larger values of $A_{\text{int}}$ as the anisotropy is increased. Hence, for the given exchange of $A_{\text{int}} = 0.6$ pJ/m, the media design indicated with the open red dot is closer to the minimum in the $\mu_0 H_{s,0}$ ($A_{\text{int}}$) curve than the media indicated by the open black dot. As a consequence, we can conclude: The increase of the anisotropy in exchange spring media will lead to the easy-to-understand increase of switching field due to higher anisotropy fields. However, since the minimum of the $\mu_0 H_{s,0}$ ($A_{\text{int}}$) curve is shifted to larger values of $A_{\text{int}}$ for larger anisotropy constants, this leads to an effect that reduces the switching field for larger anisotropy constants.

## B. Weakly coupled case ($A_{int} = 0.2$ pJ/m):

As a second example, we investigate a media design, where the switching field is a function of anisotropy in the media, shows an even more pronounced non-linear behavior, as shown in Figure 4. A kink in the $\mu_0 H_{s,0}$ ($A_{int}$) curve can be observed at about $K_1 = 1$ MJ/m³.

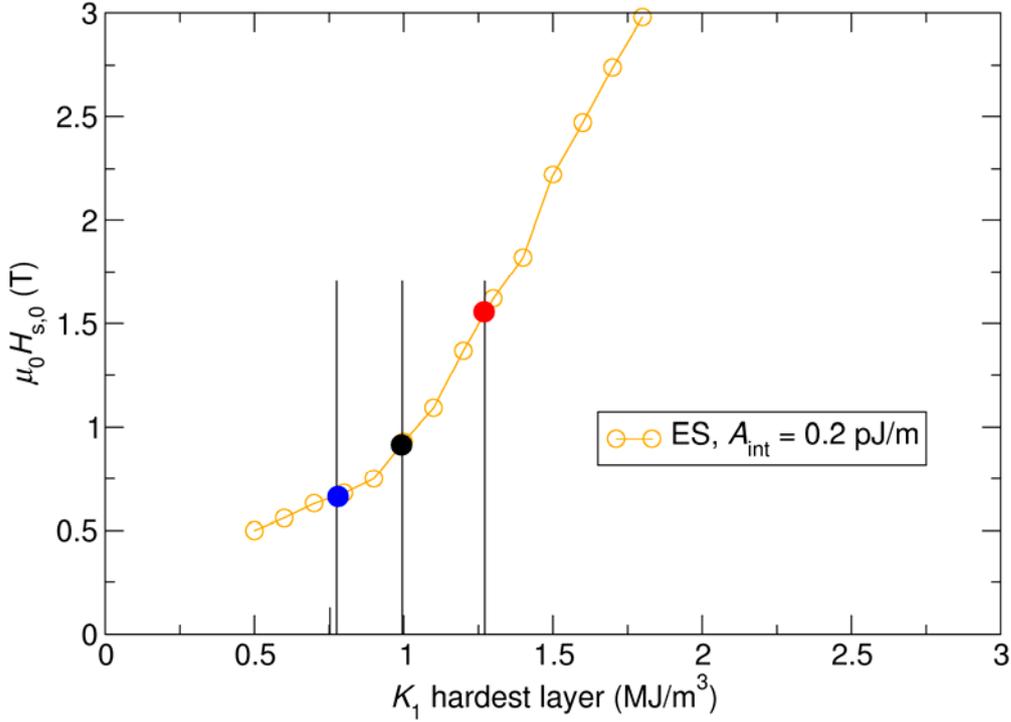

Figure 4: Detailed view of Figure 2 for $A_{int} = 0.2$ pJ/m.

By investigating the switching field as a function of the exchange coupling between the layers for three different anisotropy constants (black dot: at the kink, blue dot: left of the kink, red dot: right of the kink), the non-linear behavior can be well understood. The filled black dot indicating a media with $K_1 = 1$ MJ/m³ represents a design, where the exchange coupling of $A_{int} = 0.2$ pJ/m minimizes the switching field, as shown in Figure 3. If the anisotropy of the media is decreased to $K_1 = 0.8$ MJ/m³ (solid blue dot), the same effect as mentioned in the last paragraph can be observed, which leads only to a slight reduction of the switching field.

However, for media with anisotropy values larger than $K_1 > 1$ MJ/m³, the non-linear $\mu_0 H_{s,0}$ ($A_{int}$) curve leads to a significant enhancement of the switching field, as shown, for example, by the filled red dot in Figure 3 and Figure 4. Hence, depending on the relative strength of the anisotropy field and exchange field between the layers, the non-linear $\mu_0 H_{s,0}$ ($A_{int}$) curve may lead to an effect that further increases or decreases the switching field.

## IV. Effect of correlation of anisotropies between the soft and hard layer

In the previous chapter, simulations were performed, where the anisotropy in the soft layer was exactly 1/5 of the anisotropy of the hard layer. In order to also allow estimations for imperfectly correlated layers, numerical simulations are performed, where 1000 simulations with different realizations of the given distribution are averaged. The standard deviation of $K_1$ is assumed to be 10%.

### A. Switching field distribution for layers with correlated anisotropy

For comparison, we also perform simulations for layers with correlated anisotropy, which reproduce the results of the previous chapter for the $K_{soft}$ = 1/5 $K_{hard}$. In addition, we vary the anisotropy in the soft layer from $K_{soft}$ = 0 to $K_{soft}$ = 1/3 $K_{hard}$. For all simulations, the mean value of the anisotropy of the hardest layer is $K_1$ = 1 MJ/m³. The coercive field as a function of the exchange coupling between the soft and hard layer is shown in Figure 5 (a) for a correlated layer. The cyan curve in Figure 5 (a) agrees as expected with the black curve in Figure 3, since it is the same effect simulated with different methods.

### B. Switching field distribution for layers with uncorrelated anisotropy

In the following, we perform simulations for completely uncorrelated anisotropies in the soft layer and hard layer. The standard deviation of the normalized switching field ($\sigma \mu_0 H_{s,0} / \mu_0 H_{s,0}$) as a function of the exchange coupling between the two layers is shown in Figure 5 (c). The smallest standard deviation of the coercive field is obtained for a ratio of the anisotropy constant between the soft layer and hard layer of 1/5 and a coupling strength between the soft layer and hard layer of about $A_{int}$ = 0.35 pJ/m. Comparing Figure 5 (b) and Figure 5 (c), interestingly, the smallest standard deviation of the switching field distribution is obtained for similar values of the anisotropy constant in the correlated and uncorrelated cases.

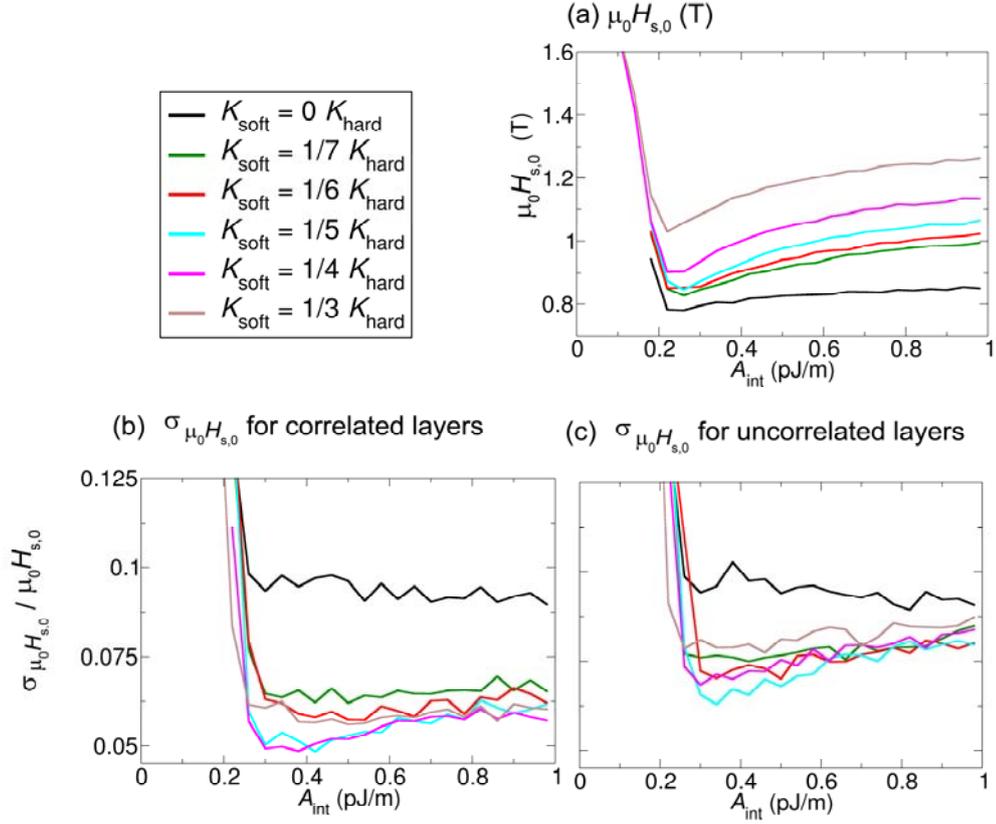

Figure 5: (a) Coercive field for soft and hard layers with correlated anisotropies. The anisotropy in the soft layer is varied. The smallest coercive field is obtained for a perfectly soft layer and coupling of $A_{int}$ = 0.25 pJ/m. (b) Normalized standard deviation of the coercive field as a function of the exchange coupling between the soft and the hard layer for layers with correlated $K_1$. (c) Uncorrelated $K_1$. The mean value of the anisotropy of the hard layer is $K_1$ = 1 mJ/m³.

*C. Comparison layers with correlated and uncorrelated anisotropies*

The standard deviation of the coercive field as a function of the anisotropy in the soft layer for the correlated case and uncorrelated case is compared in Figure 6. The smallest standard deviation is obtained for both cases for a value of the anisotropy constant of $K_{1,soft}$ = 1/5 x $K_{1,hard}$. For the simulations of Figure 6, the exchange coupling between the soft and hard layer is $A_{int}$ = 0.4 pJ/m. The smallest value of the standard deviation is obtained for the case of uncorrelated anisotropies in the layers. A correlated and uncorrelated variation of the anisotropy constant in the soft layer of 10% leads to a change of the coercive field of only about 6.5% and 4.8% for the optimized exchange spring structure, respectively.

Interestingly, if the anisotropy in the soft layer is optimized as in Ref [1] or the exchange is optimized as in this work, the switching field distribution can be reduced independently by about 35%.

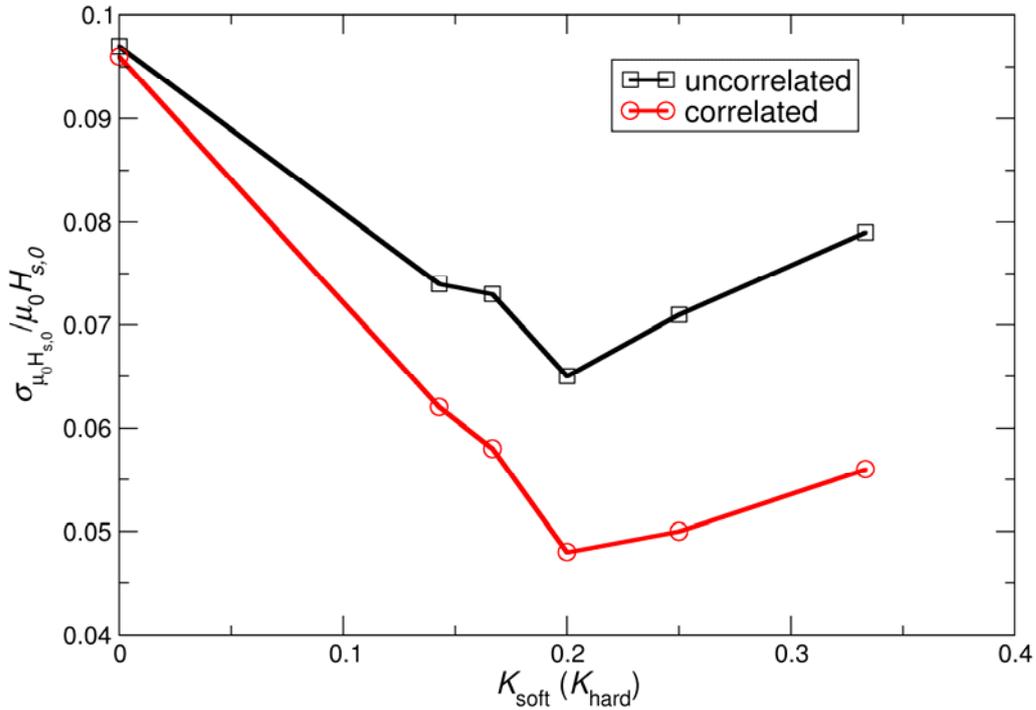

Figure 6: Standard deviation of the normalized switching field as a function of the anisotropy of the soft layer for $A_{int}$ = 0.4 pJ/m.

### V. Fundamental switching field distribution due to thermal fluctuations

Besides the switching field distributions due to variations of the anisotropy values in the layers, there exists an intrinsic switching field distribution due to thermal fluctuations. This distribution occurs even for perfect grains of the same size and without any material parameter distributions and stray field interactions. This intrinsic switching field distribution can be regarded as a fundamental limiting factor for magnetic recording.

In order to simulate this intrinsic distribution, the switching probability (*P*) as a function of a field pulse strength ($\mu_0 H$ (T)) is calculated for the single-phase media and the exchange spring structure of the previous section. The temperature is *T* = 300 K. Thermal activation is introduced in the Landau–Lifshitz equation by a stochastic thermal field, which is added to the effective field [10–12]. We numerically integrate the stochastic Landau-Lifshitz-Gilbert equation of motion using a semi-implicit method [13]. The same field pulse as in the previous section is used. The field is ramped to the maximum field linearly within a time of 0.1 ns. For 1 ns, the field is applied with a magnitude *H*. After this time, the field is decayed to zero within 0.1 ns. In order to obtain a good statistic, each grain is simulated 128 times for the same field pulse with a different seed of the random number generator for the stochastic random field. Out of these 128 simulations for one field pulse, the switching probability is calculated. Figure 7 shows the simulated switching probability as a function of field pulse strength *H* for the exchange spring media and the single-phase media.

In order to obtain a better understanding of the simulated switching field distributions in the following, an analytic estimate of the switching probability as a function of field pulse strength is derived. We follow the arguments of Ref [14] in order to derive the probability *P* that the particle switches, if an external field pulse of duration $t_{\text{pulse}}$ and strength *H* is applied opposing the initial magnetization. The probability of switching *P* as a function of field pulse strength *H* is given by:

$$P(H) = 1 - \exp\left[-\frac{t_{\text{pulse}}}{\tau(H)}\right] \qquad (8)$$

where

$$\tau(H) = \frac{1}{f_0} \exp\left[\Delta E_0 \left(1 - \frac{H}{H_0}\right)^n\right]. \qquad (9)$$

Since, in the numerical simulation, the external field is applied at a finite angle with respect to the easy axis, we assume that *n* = 1.5, which is supposed to be a good approximation for the exponent [15, 16]. The analytic switching probability as a function according to Eq. (8) is plotted in Figure 7. The coercive field at zero temperature $\mu_0 H_{\text{s},0}$ and the energy barrier $\Delta E$ are obtained from independent micromagnetic simulations. The energy barrier is calculated using the nudged elastic band method [17]. The obtained parameters for $\mu_0 H_{s,0}$ and $\Delta E$ are summarized in Table 2. The only free parameter is the attempt frequency $f_0$, which is used as a fitting parameter to obtain the best agreement with the simulated results. Interestingly, the numerically obtained curves can be fitted excellently with the analytic theory according to Eq. (8), if, for both structures (exchange spring media and single-phase media), basically the same attempt frequency of $f_0 \sim 30$ GHz is assumed. It is worth noting that with this single fitting parameter, not only does the switching field at $T=300$ K ($\mu_0 H_{s,300K}$) agree very well but also the slope of the $P(H)$ curve agrees excellently.

Analyzing the analytical formula of Eq. (8), as it was done in Ref [14], an inherent advantage of the exchange spring structure appears. As it can be seen in Ref [14], the standard deviation of the switching field $\sigma_c$ decreases with increasing energy barrier of the invested structure. Hence, due to the larger energy barrier of the exchange spring structure compared to the single-phase media, the switching field distribution $\sigma_{\mu_0 H_{s,300K}}$ is expected to decrease.

In order to extract the thermal switching field distribution $\sigma_{\mu_0 H_{s,300K}}$ from the numerical simulated data, we assume that the derivative of the switching field follows a Gaussian function:

$$\frac{dP(\bar{H})}{d\bar{H}} = \frac{1}{\sqrt{2\pi}\sigma_{\mu_0 H_{s,300K}}} \exp\left[-\frac{1}{2}\left(\frac{\bar{H}-H_s}{\sigma_{\mu_0 H_{s,300K}}}\right)^2\right]. \tag{10}$$

The switching probability as a function of field pulse strength is obtained by integration as follows:

$$P(H) = \int_{-\infty}^{H} \frac{dP(\bar{H})}{d\bar{H}} d\bar{H} = \frac{1}{2} \text{erf}\left[\frac{\mu_0(H-H_s)}{\sqrt{2}\sigma_{\mu_0 H_{s,300K}}}\right] + \frac{1}{2}. \tag{11}$$

From the fit of Eq. (11) with the numerical data, we obtain $\sigma_{\mu_0 H_{s,300K}}$ and $H_s$, which are summarized in Table 2.

Importantly, it can be seen that indeed, the $\sigma_{\mu_0 H_{s,300k}}$ is significantly smaller by about 33% in the case of the exchange spring media. This is also predicted by the analytical estimate according to Eq. (8), since the slope of the $P(H)$ loop of the Gaussian fit and the analytical estimate agree excellently. The origin is the different energy barriers of the two structures.

It is worth noting that the non-Gaussian fit according to Eq. (8) is better than the fit assuming a Gaussian distribution using Eq. (11). Hence, the thermally introduced switching field distribution $\frac{dP(\bar{H})}{d\bar{H}}$ is not exactly a Gaussian function.

By comparing the obtained value of the thermal switching field distributions (Table 2) with the switching field distribution due to variations of the intrinsic anisotropy constant (Table 1), it can be seen that they are comparable in size. Hence, the thermal induced switching field distributions are an important factor, and even if $\sigma_K = 0$, thermal fluctuations contribute in the case of the single-phase media, with about 3% to $H_{s,300K}$ distributions.

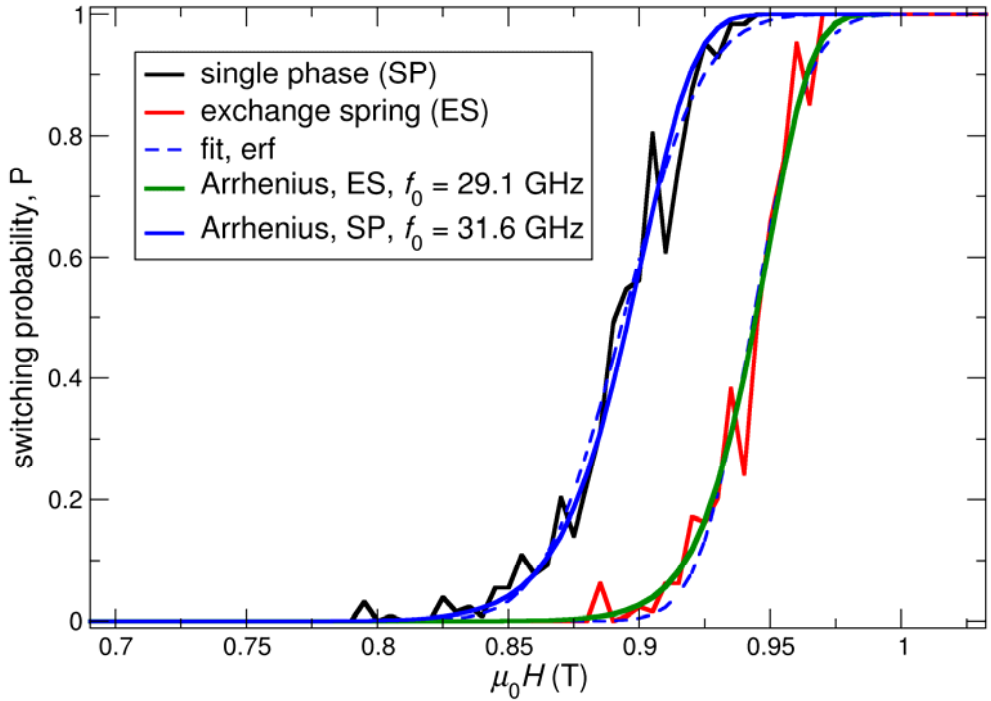

Figure 7: Switching field distribution due to thermal fluctuations for an exchange spring media and single-phase media at $T = 300$ K. The material parameters of the media are shown in Table 2.

Table 2: Thermal switching field distributions due to simulations at $T = 300$ K for grains without any intrinsic material parameter distributions.

|  | Single-Phase | Exchange Spring |
|---|---|---|
| **Input:** |  |  |
| $d$ (nm) | 12 | 12 |
| $t$ (nm) | 10.4 | 10.4 |
| $K_{max}$ (MJ/m³) | 0.46 | 1.0 |
| $J_s$ (T) | 0.7 | 0.7 |
| $A_{inter}$ (pJ/m) | 1000 | 60 |
| **Output:** |  |  |
| $\mu_0 H_{s,0}$ (T) | 0.99 | 1.03 |
| $\mu_0 H_{s,300K}$ (T) | 0.89 | 0.94 |
| $\Delta E$ ($k_B T_{300}$) | 113.8 | 155.9 |
| $\sigma_{\mu_0 H_s, 300K}$ (T) | 0.024 | 0.016 |
| $\sigma_{\mu_0 H_s, 300K} / \mu_0 H_{s,300K}$ | 2.7% | 1.7% |

## VI. Implication of switching field distribution for bit-patterned and granular media

### A. Bit error rate for bit-patterned media

In this section, we discuss the implication of the reduced switching field distribution on the expected bit error rate (BER) of BPM. In order to derive the BER for BPM, let us assume a typical phase diagram of a bit-patterned media island as a function of downtrack position and anisotropy constant. The recording head is moved across one BPM island, and the polarity of the head is reversed from "down" to "up" and finally to "down" again. Initially, the BPM islands points in the "down" direction, which is denoted by white in the phase plot in Figure 8. If the island is reversed to "up," the state is denoted in Figure 8 as black. Depending on the anisotropy constant of the island and the downtrack position of the island, the island might be able to be reversed to the "up" state. As it can be seen in Figure 8, there is some certain interval in anisotropy constant that leads to a successful writing of the "up" state. Hence, if the distribution of the anisotropy constant is sufficiently small, there is no significant error in writing. In more detail, the BER of BPM can be calculated by assuming that the anisotropy constants are distributed according a Gaussian function, as indicated by the blue Gaussian in Figure 8. Hence, if the tail of the Gaussian function spreads out in the region, where the islands can not be written any more, significant errors are introduced.

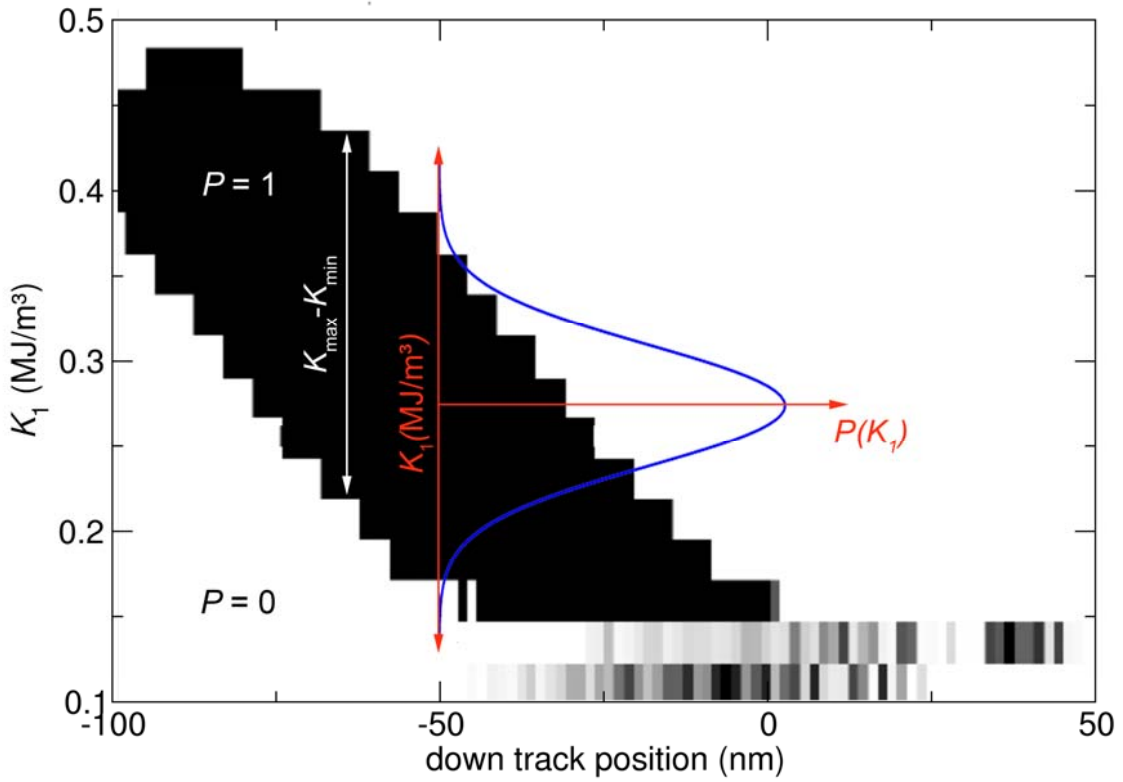

Figure 8: Phase plot of successful switching and non-switching for a bit-patterned media element as a function of anisotropy constant and downtrack position.

The BER can be calculated by integrating over the region of anisotropy constants, where successful writing is possible, and weighting it with the probability that this particular anisotropy occurs according to the Gaussian distribution:

$$BER = 1 - \int_{-K_{min}}^{K_{max}} \frac{1}{\sigma_K \sqrt{2\pi}} e^{-\frac{1}{2}\left(\frac{K-K_{aver}}{\sigma_K}\right)^2} dK, \qquad (12)$$

where $K_{max}$ and $K_{min}$ are the maximum anisotropy and minimum anisotropy that can be successfully written with a particular field pulse, respectively. $\sigma_K$ is the standard deviation of the anisotropy constant, and $K_{aver}$ is the average anisotropy constant. The highest BER is obtained if the media is designed such that the $K_{aver} = 1/2(K_{max} + K_{min})$, which is assumed to be the case in the following. Next, we transform Eq. (12) to dimensionless variables by:

$$\begin{aligned} x &= (K - K_{aver})/K_{aver} \\ \Delta x &= \frac{K_{max} - K_{min}}{2 K_{aver}} \\ \tilde{\sigma}_K &= \sigma_K / K_{aver} \end{aligned} \quad (13)$$

leading to:

$$BER = 1 - \int_{-\Delta x}^{\Delta x} \frac{1}{\tilde{\sigma}_K \sqrt{2\pi}} e^{-\frac{1}{2}\left(\frac{x}{\tilde{\sigma}_K}\right)^2} dx \quad (14)$$

In the following, we assume that the minimum and maximum anisotropy is assigned to values so that the BER is $10^{-3}$. This will be our reference structure. Under the assumption of $\tilde{\sigma}_K = 0.03$, a BER of $10^{-3}$ is obtained for $\Delta x = 0.0987$:

$$BER = 1 - \int_{-\Delta x}^{\Delta x} f(x, \tilde{\sigma}_K) dx = 10^{-3}. \quad (15)$$

In the case of the exchange spring media at the optimized design point (for $A_{int}$=0.6 pJ/m), the ratio of the standard deviation of the anisotropy constant of the exchange spring media and single-phase media is (using the values of Table 1) $\sigma_{K,ES}/\sigma_{K,SP} = 1.52$. Hence, compared to single-phase media, the interval of the integration can be extended by a factor of 1.56 for the exchange spring media:

$$BER = 1 - \int_{-1.56\Delta x}^{1.56\Delta x} f(x, \tilde{\sigma}_K) dx = 2.8 \times 10^{-7}, \quad (16)$$

since a 1.56-times higher change in anisotropy constant in exchange spring media still leads to correct reversal of the grain. It can be seen that due to the improved switching field distribution in exchange spring media, a tremendous improvement of the BER can be expected. In this estimate, the stray field interaction between the islands was not taken into account. As published elsewhere, the stray field interaction between islands can be regarded as an additional contribution to the switching field of about $\sigma_{stray} = 0.03$T. If this additional distribution is taken into account, which is assumed to be the same for exchange spring and single-phase media, an improvement of *BER* by a factor of about 8 is still expected for exchange spring media, as shown in Table 3.

Table 3: Total switching field distributions and effect on BER if the stray field interaction is considered ($\sigma_{stray}$ = 0.03 T) . Data are given relative to a standard media, where it is assumed that a switching field distribution of 3% leads to a BER of $10^{-3}$.

|  | $\sigma_K$ | Single-Phase | Exchange Spring |
|---|---|---|---|
| $\sigma_{tot,nostray}$ (T)=($\sigma_{\mu 0Hs,300K}^2 + \sigma_{\mu 0Hs,300K}^2$)$^{0.5}$ | 3% | 0.039 | 0.025 |
| BER for $\sigma_{tot,nostray}$ | 3% | $1.1 \times 10^{-2}$ | $7.8 \times 10^{-5}$ |
| $\sigma_x$ for $\sigma_{tot,nostray}$ (nm) | 3% | 0.78 | 0.5 |
| $\sigma_{tot, stray}$ (T)=($\sigma_{\mu 0Hs,300K}^2 + \sigma_{\mu 0Hs,300K}^2 + \sigma_{stray}^2$)$^{0.5}$ | 3% | 0.049 | 0.039 |
| BER for $\sigma_{tot, stray}$ | 3% | $4.3 \times 10^{-2}$ | $1.1 \times 10^{-2}$ |
| $\sigma_x$ for $\sigma_{tot, stray}$ (nm) | 3% | 0.98 | 0.78 |
| $\sigma_{tot, stray}$ (T)=($\sigma_{\mu 0Hs,300K}^2 + \sigma_{\mu 0Hs,300K}^2 + \sigma_{stray}^2$)$^{0.5}$ | 5% | 0.063 | 0.048 |
| BER for $\sigma_{tot, stray}$ | 5% | $1.1 \times 10^{-1}$ | $3.9 \times 10^{-2}$ |
| $\sigma_x$ for $\sigma_{tot, stray}$ (nm) | 5% | 1.26 | 0.96 |

*A. Jitter for granular media*

Besides the BER for bit-patterned media, we can also estimate the jitter in granular recording. In the following, we assume a head field gradient of 0.05 T/nm. The jitter is estimated as the product of $\sigma_{tot,no\,stray}$ divided by the head field gradient. The results are summarized in Table 3. Interestingly, using the data of Ref [18], the fundamental thermal jitter for heat-assisted magnetic recording using constant laser heating, and an external field of $\mu_0 H_{head}$ = 0.8 T, one obtains $\sigma_{x,heat\,assist}$ = 0.93 nm, which is significantly larger than for non-heat-assisted recording on graded media. The estimate for heat-assisted recording was done without taking into account the stray field interaction. However, stray field interaction plays a smaller role in heat-assisted recording and might be neglected for a good approximation. It also has to be noted that in this study, the investigated grain sizes are larger than in Ref [18] . However, the jitter due to variation of the intrinsic anisotropy distribution does not depend on the grain size. If we just compare these grain sizes independently of jitter value for heat-assisted recording, assuming a 3% distribution of $T_c$ and temperature gradient of 20 K/nm, and for exchange spring media a 5% distribution of $K_1$ values and a head field gradient of 0.05 T/nm, we obtain for heat-assisted recording $\sigma_{x,\text{heat assist}} = 0.9$ nm and exchange spring structures $\sigma_{x,\text{exchange spring}} = 0.69$ nm. For comparison for the single-phase media, one obtains $\sigma_{x,\text{exchange spring}} = 1.0$ nm. Hence, even so, it is often argued that the effective head field gradient in heat-assisted recording is significantly higher than in conventional recording, taking into account realistic distributions, that one sees that this does not result in significantly better jitter values compared to optimized exchange spring structures. In this case, conventional recording using exchange spring structure leads to even better jitter values.

## VII.
## Conclusion

We have shown that BER in BPM can be drastically improved by an exchange spring media design. The origin of the improved BER can be found in a weak dependence of the switching field as a function of the anisotropy of both layers. This effect can be explained by the non-linear relation of the switching field as a function of the exchange coupling between the soft and the hard layer. Independently, if correlated or uncorrelated distributions of the anisotropy constant in the soft and hard layer are assumed, the minimum switching field distribution is obtained if the soft layer anisotropy is about 1/5 of the hard layer anisotropy. The switching field distribution depends significantly on the exchange coupling between the soft and the hard layer.

As a second effect, it was shown that thermal fluctuations during recording at $T = 300$ K lead to significant additional transition jitter. This jitter increases as the thermal stability of the media decreases, leading to particular challenges for small grains at ultra-high recording density. Due to the improved thermal stability of exchange spring structure, this contribution is significantly decreased compared to single-phase media.

A simple analytical estimate is given in order to predict which improvement of BER can be expected due to the improved switching field distribution. In exchange spring media, the BER is expected to be improved by at least one order of magnitude.

The jitter contribution that is obtained due to a variation of 5% in anisotropy constant in the optimized exchange spring structure is as small as $\sigma_{x,\text{exchange spring}} = 0.69$ nm. For comparison an optimized heat-assisted recording media assuming 3% distribution of $T_c$ and very narrow heat sport with a the full width at half maximum of the spatial Gaussian of the temperature profile of 20nm leads to a cross track jitter of $\sigma_{y,\text{heat-assists}} = 0.9$ nm and a down track jitter of $\sigma_{y,\text{heat-assists}} = 0.5$ nm.

Hence, the overall jitter is similar for the presented exchange spring media and the optimized heat-assisted recording media.

If we apply the rule of thumb that the jitter is about 20% of the minimum bit length, we obtain a bit length of 3.9 nm for the optimized exchange spring with a jitter value of 0.78 nm. Here, further studies have to be performed how smaller grain diameters influence the studies performed in this paper. Under the assumption of a bit aspect ratio of 1:3.1, the areal density can be extrapolated to be around 13 TBit/inch², which is indeed very close to the original proposed areal density of 10 TBit/inch² in 2004 for exchange spring media [3].

The financial support by the Austrian Federal Ministry of Science, Research, and Economy and the National Foundation for Research, Technology and Development, the SFB project F4112-N13, and the Advanced Storage Technology Consortium (ASTC) is gratefully acknowledged. The computational results presented have been achieved using the Vienna Scientific Cluster (VSC).